\documentclass{ctr}

\usepackage{ctrfont}
\usepackage{natbib}

\usepackage{url}
\usepackage{amsmath}
\usepackage{amssymb}

\usepackage{graphicx}
\usepackage{graphicx}
\usepackage{subcaption}

\title{Quantifying structural uncertainty in the BHR model for variable-density flows}
\shorttitle{Quantifying structural uncertainty in the BHR model}
\author{Z. Huang, J. Hayes \and G. Iaccarino}
\shortauthor{Huang,  Hayes  \&  Iaccarino }

\begin{document}


\maketitle

\section{Motivation and objectives} 

Variable-density flows play an important role in a variety of physical phenomena, including inertial confinement fusion \citep {Lindl1992}, oceanic flows \citep {Boyd}, and stellar evolution and the explosion of supernovae \citep {Zingale2009} with a wide range of length scales. Variable-density effects arise whenever two fluids of differing molecular weights begin to mix or when thermal or compressibility effects significantly modify the fluid density. When the fluid densities are close to each other, i.e., the density ratio is close to 1,  the density field is usually modeled as a passive scalar with the Boussinesq approximation. For a flow with sufficiently high fluid density variance, which is characterized by Atwood number ($At = (\rho_2 - \rho_1)/(\rho_2 + \rho_1)$), where $\rho_1$  and $\rho_2$ are light and heavy fluid density, respectively, the density gradients play a very important role in the transport and evolution of the turbulent characteristics of the flow. Motivated by these applications, significant progress has been made in studying the variable-density flows using direct numerical simulations (DNS) and large-eddy simulations (LES). However, these approaches remain computationally expensive, especially for complex applications. For complex applications, computationally tractable and efficient turbulence models based on transport equations, which predict the “averaged”  behavior of the turbulent mixing zone, are employed. Due to ensemble averaging of the second- and higher-order correlations of turbulent fluctuations, additional terms arise and should be modeled. 

Among the models of variable-density flow, the Besnard-Harlow-Rauenzahn (BHR) model \citep {Besnard1992} is very important and computationally tractable. This model is based on the evolution equations arising from second-order correlations and gradient-diffusion approximations. Using a mass-weighted averaged decomposition, the original BHR model includes full transport equations for the Reynolds stresses, turbulent mass-flux velocity, density fluctuations, and turbulence kinetic energy dissipation rate. Many advanced variants \citep{Schwarzkopf2011, Schwarzkopf2016} of the BHR model have been developed to predict the variable-density flows. 

The Reynolds-averaged Navier-Stokes (RANS) models, such as $k-\epsilon$ and $k-\omega-SST$, are computationally efficient and widely employed in simulating turbulent flows in complex engineering applications. However, the simplifications and assumptions  made in RANS models induce a significant degree of epistemic uncertainty in numerical predictions. The structural uncertainty in RANS models has been quantified by incorporating the perturbations to the eigenvalues  \citep [] { Gorle2013, Emory2013} and the eigenvectors \citep [] { Mishra2016, Thompson2019} of the modeled Reynolds stress tensor. In this methodology, the structural uncertainties in turbulence models are quantified by injecting uncertainty into the shape and orientation of the Reynolds stress tensor to quantify the variability and biases. Specifically,  the perturbations of the Reynolds stress anisotropy are governed by sampling from the extreme states of the turbulence componentiality, and the perturbations to the Reynolds stress eigenvectors are guided by the maximal states of the production mechanism. Consequently, the approximate bounds of the structural uncertainty can be estimated within five RANS simulations \citep [] {Iaccarino2017}. The eddy-viscosity hypothesis  assumes that the Reynolds stress has the same eigenvectors as the mean rate of strain, which is known to be invalid in complex flows such as flow with curvy boundaries and separations. The physical complexity is further increased by the buoyancy effect. Currently, the structural uncertainty framework accounts for the discrepancy between experimental observations and high-fidelity simulations for many flows with separations \citep [] {Iaccarino2017} and curvy boundaries \citep [] { Thompson2019}. However, the extension to variable-density flow is very limited due to the significantly increased complexity, for example, the buoyancy effect in the turbulent closures and transport equations for turbulent mass-flux velocities.

In this brief, we try to develop a comprehensive framework to identify, quantify, isolate, and reduce the uncertainties in the original BHR model \citep {Besnard1992} for variable-density flows. Because the eigenspace perturbation of Reynolds stress successfully accounts for the structural uncertainty in many flows, we first extend this methodology to the BHR model for variable-density flows with different $At$. The philosophy of eigenspace perturbation in Reynolds stress, i.e., quantifying the states of maximizing and minimizing the production mechanism, has led to the proposals of the structural uncertainty quantification strategy in the BHR model for variable-density flows. Several canonical flows, including one-dimensional (1D) Rayleigh-Taylor instability (1DRT), Rayleigh-Taylor mixing in a tilted rig (TR) and turbulent jet mixing flow, are investigated.


\section{Mathematical formulations }
In this section, we introduce the basic knowledge of the original BHR model for variable-density flows \citet {Besnard1992} and the eigenspace perturbation framework \citet { Mishra2017}, respectively. The BHR model and the eigenspace perturbation method are implemented in OpenFOAM to simulate the variable-density flows.

\subsection{BHR model for variable-density flow}
For the Reynolds decompositions of $\rho = \overline{\rho} + \rho'$, $u_i = \overline{u_i} + u_i'$ and $p = \overline{p} + p'$ for density, velocity, and pressure, respectively, where the over-bar denotes the uniformly weighted ensemble. However, for variable-density flows, a mass-weighted averaging called Favre decomposition is employed 
\begin{equation}
u_i=\widetilde{u_i} + u_i'',        \label{2}
\end{equation}
where $\widetilde{u_i} = \overline{\rho u_i}/\overline{\rho}$ and $u_i''$ represents the mass-weighted fluctuation with $\overline{\rho u_i''} = 0$. Then, the mass-weighted average velocity $\widetilde{u_i}$ can be rewritten by applying the Reynolds decomposition
\begin{equation}
\tilde {u_i} = \frac{\overline{\rho u_i}} {\overline {\rho}}  = \overline{u_i} +    \frac{\overline{\rho' u_i'}} {\overline {\rho}},   \label{3}
\end{equation}
where $ \overline{\rho' u_i'}/ \overline {\rho}$ is the turbulent mass-flux velocity, which is usually denoted as $a_i$ . Another important quantity in the BHR model equations is the density self-correlation, $b$, defined as
\begin{equation}
  b = - \overline{ \rho ' \left(\frac{1} {\rho}\right)'  },   \label{4}
\end{equation}
which can also be written as $b = \overline{\rho'^2/\rho \overline{\rho}}$. The generalized Reynolds stress tensor,  $R_{ij} = \overline{\rho u_i'' u_j''}$, is obtained by Favre decomposition. 

The resulting modeled governing equations are given by   
\begin{equation}
 \frac{\partial \overline{\rho}}{\partial t}  +   \frac{\partial \overline{\rho} \widetilde{u_j}}{\partial x_j}  = 0, \label{5}
\end{equation}

\begin{equation}
 \frac{\partial \overline{\rho}  \widetilde{c}_{\alpha}}{\partial t}  +    \frac{\partial \overline{\rho} \widetilde{u_j} \widetilde{c}_{\alpha} }{\partial x_j}  =     \frac{\partial  }{\partial x_j}  \left( \frac{ \mu_T \partial \widetilde{c}_{\alpha}  }{ \sigma_c  \partial x_j}  \right),      \label{6}
\end{equation}

\begin{equation}
 \frac{\partial \overline{\rho} \widetilde{u_i} }{\partial t}  +   \frac{\partial \overline{\rho} \widetilde{u_j} \widetilde{u_i}}{\partial x_j}  = -\frac{\partial (\overline{p} \delta_{ij} + R_{ij} )}{\partial x_j}   + \overline{\rho} g_i,   \label{7}
\end{equation}


\begin{equation}
 \frac{\partial \overline{\rho} k }{\partial t}  +   \frac{\partial \overline{\rho} k \widetilde{u_i}}{\partial x_j}  = a_j \frac{\partial \overline{p} }{\partial x_j}   -   R_{ij} \frac{\partial \widetilde{u_i}}{\partial x_j}   +  \frac{\partial  }{\partial x_j}  \left( \frac{ \mu_T \partial  k  }{ \sigma_k  \partial x_j} \right)      - \overline{\rho} \frac{k^{3/2}}{S},  \label{9}
\end{equation}

\begin{equation}
\begin{split}
 \frac{\partial \overline{\rho} S }{\partial t}  +   \frac{\partial \overline{\rho} S \widetilde{u_i}}{\partial x_j}  = & \frac{S}{k} \left[ \left( \frac{3}{2} - C_4  \right)  a_j \frac{\partial \overline{p} }{\partial x_j} -   \left( \frac{3}{2} - C_1  \right) R_{ij} \frac{\partial \widetilde{u_i}}{\partial x_j}  \right]   \\
 &+ \frac{\partial  }{\partial x_j}  \left( \frac{ \mu_T \partial  S  }{ \sigma_S  \partial x_j} \right)     -   \left( \frac{3}{2} - C_2  \right)  \overline{\rho} k^{1/2},  \label{10}
\end{split}
\end{equation}

\begin{equation}
\begin{split}
 \frac{\partial \overline{\rho} {a_i} }{\partial t}  +   \frac{\partial \overline{\rho} \widetilde{u_j} {a_i}}{\partial x_j}  = & b \frac{\partial \overline{p} }{\partial x_i}   -  \overline{\rho} a_j \frac{\partial (\widetilde{u_i} - a_i)}{\partial x_j}   -   \frac{R_{ij}}{\overline{\rho}}\frac{\partial \overline{\rho}}{\partial x_j}   \\
&+ \overline{\rho} \frac{\partial a_i a_j}{\partial x_j}+  \frac{\partial  }{\partial x_j}  \left( \frac{ \mu_T \partial  a_i  }{ \sigma_a  \partial x_j} \right) -
C_a \overline{\rho} a_i \frac{k^{1/2}}{S},    \label{11}
\end{split}
\end{equation}

\begin{equation}
 b = \frac{\alpha_1 \alpha_2 \left(\rho_1 - \rho_2\right)}{\rho_1\rho_2}.  \label{12}
\end{equation}
where $c_{\alpha}$ is the mass fraction, $\mu_T$ is the turbulent eddy viscosity and is modeled as  $\mu_T = C_{\mu} \overline {\rho} S k^{1/2} $, $R_{ij} = 2/3\overline{\rho}k \delta_{ij} - 2 \mu_T S_{ij} $ is the generalized Reynolds stress with the rate of mean strain definition $ S_{ij} = 1/2 \left[   {\partial \widetilde{u_i}}/{\partial x_j}+{\partial \widetilde{u_j}}/{\partial x_i}  - {2}/{3} {\partial \widetilde{u_k}}/{\partial x_k}  \delta_{ij}  \right]$, $k$ is the turbulent kinetic energy,  $S$ is the turbulent mixing length, and $a_i$ and $b$ are the turbulent mass-flux velocity and the density self-correlation, respectively, as discussed above. The averaged density $\overline{\rho} = \alpha_1 \rho_1 + \alpha_2 \rho_2$, and $ \alpha_1 + \alpha_2 = 1$. According to previous studies \citep {Schwarzkopf2011, Schwarzkopf2016}, the first two terms $G$ and $P$ in the right handside of turbulent kinetic energy transport equation, are the production,  where $G = a_j {\partial \overline{p} }/{\partial x_j} =  \bold{a} \cdot  \bold{p}_g$ where $ \bold{a}$ is the turbulent mass-flux velocity vector and $\bold{p}_g$  is the pressure gradient, and $  P =  -R_{ij} {\partial \widetilde{u_i}}/{\partial x_j} $.

\subsection{Eigenspace perturbations}
The methodology to perturb the eigenvectors of the Reynolds stress tensor in the RANS modeling framework without using any additional data or modeling assumptions has been proposed \citep [] {Gorle2013} and developed for many engineering flows \citep []{Emory2013, Gorle2015, Mishra2016, Mishra2017, Thompson2019}.  The structural uncertainty framework works very well for many flows with separations and enables us to account for the discrepancy between experimental observations and high-fidelity simulations. The basic methodology is described as follows. The turbulence anisotropy tensor is
\begin{equation}
a_{ij} = \frac{R_{ij}}{2 \overline{\rho}k} - \frac{1}{3} \delta_{ij},  \label{13}
\end{equation}
The condition of realizability and the Cauchy-Schwartz inequality require that $ -1/3 \le a_{ij} \le 2/3$ for $i = j$  and $ -1/2 \le a_{ij} \le 1/2$ for $i \ne j$.    The anisotropy tensor can be represented by an eigenvalue decomposition 
\begin{equation}
a_{ij} =  v_{in} \Lambda_{nl} v_{lj}, \label{13}
\end{equation}
where $v$ denotes the eigenvector matrix and $\Lambda$ denotes the eigenvalues, and they are ordered by $\lambda_1 \ge \lambda_2  \ge \lambda_3 $ without any loss of generality. Therefore, the Reynolds stress can be expressed as  
\begin{equation}
R_{ij} = 2 \overline{\rho} k  \left( v_{in} \Lambda_{nl} v_{lj} + \frac{1}{3} \delta_{ij} \right),  \label{13}
\end{equation}
Perturbing the eigenvalues and eigenvector by injecting the uncertainties, one obtains the perturbed Reynolds stress.
\begin{equation}
R_{ij} = 2 \overline{\rho} k  \left( v_{in}^* \Lambda_{nl}^* v_{lj}^* + \frac{1}{3} \delta_{ij} \right),  \label{13}
\end{equation}

\subsection{Structural uncertainty quantification in the BHR model}

 The purpose of this investigation is to develop a framework to identify and quantify the uncertainties of the BHR model for variable-density flows.  For eigenspace perturbation in Reynolds stress, the states of maximizing and minimizing the production mechanism are very important. Similarly, we propose a perturbation strategy to estimate the maximal and minimal states of the production to quantify the structural uncertainty in the BHR model for variable-density flows. Note that the term $G$ in the production is the inner product of the turbulent mass-flux velocity vector and mean pressure gradient; the alignment of the two vectors gives the maximum and minimum. Recalling the maximal and minimal states of $P$ arising from the eigenspace perturbation, we can capture the maximal and minimal states of the term $P$ in the BHR model. Therefore, we can predict the model structural uncertainty intervals quantified by the production mechanism.

By perturbing the orientation of $\bold{a}$,  i.e.,  the angle between the $\bold{a}$ and $\bold{p}_g$, the extreme states of $G$ are very easily obtained as $G_0$ and $G_{\pi}$, which correspond to the angle between the two vectors, $0$ and $\pi$, respectively.  Perturbing the eigenvalue and eigenvectors in $P$ gives the five extreme states. By combining them, one can quantify the extreme states of the production in the BHR model by the combination of $G$ and $P$ perturbation, i.e., with the 10 states.

The BHR model, the eigenspace perturbation method, and the perturbation of $G$ are implemented in OpenFOAM to simulate variable-density flows. In this brief, we investigate three variable-density flows: 1DRT, two-dimensional Rayleigh-Taylor-driven mixing in a TR (2DTR), and turbulent jet mixing. 

\section{Results and discussions}

We use the structural uncertainty quantification method by maximizing and minimizing the production mechanism in the BHR model for variable-density flows. The eigenspace perturbations in $P$ are performed to study the partial structural uncertainty. The uncertainty intervals caused by $G$ are compared with the eigenspace perturbations in $P$. Finally, the structural uncertainties of the BHR model are quantified by the combination of the perturbations of $G$ and $P$. The numerical results show that the 10 extreme cases enable us to account for the discrepancy between experimental observations and high-fidelity simulations.

\subsection{1DRT}

We first consider the 1DRT case at a low Atwood number, $At = 0.04$, due to the available air-helium gas channel experimental measurements \citep [] { Banerjee2010a} and numerical simulations \citep [] { Banerjee2010b}. Furthermore, the turbulent mass-flux velocity and pressure gradient have the same direction due to the 1D setup. Therefore, $G$ is certainty, and no perturbation is needed. The production perturbation is simplified to the  eigenspace perturbation  \citep [] {Emory2013, Iaccarino2017} in the term of $P$ with the five extreme cases. 

\begin{figure}
\begin{center}
\includegraphics[width=1\textwidth]{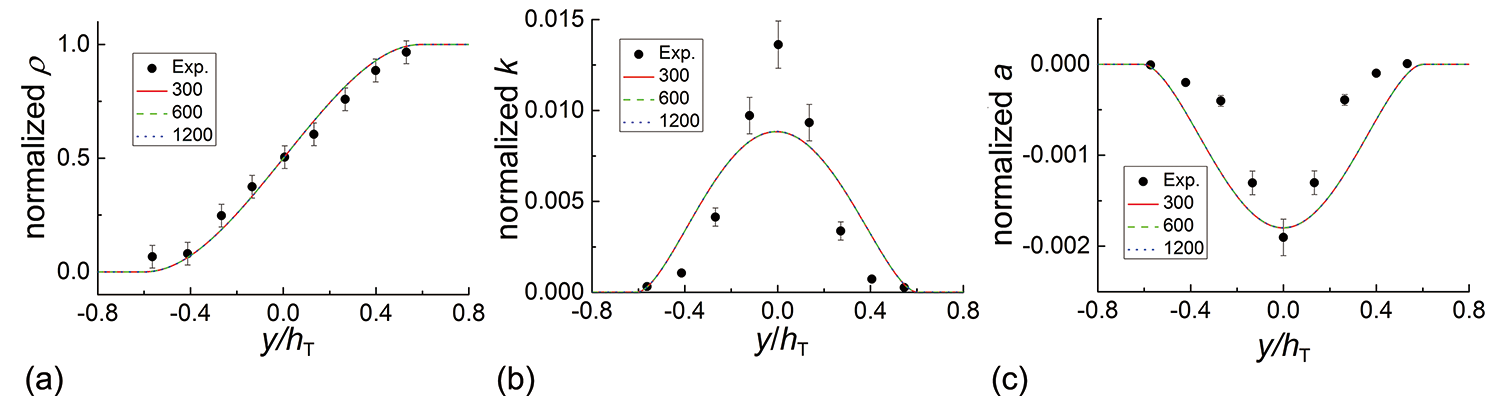}
\caption{Simulations of 1DRT by the BHR model compared with the experimental measurements by \citet { Banerjee2010a} for (a) normalized density, (b) normalized turbulent kinetic energy, and (c) normalized turbulent mass-flux velocity.}
\label{fig1}
\end{center}
\end{figure}

In the experiment by \citet [] {Banerjee2010a}, two gas streams, one containing air and the other containing a helium-air mixture, flow in parallel with each other separated by a thin splitter plate. The two streams meet at the end of the splitter plate, leading to the formation of an unstable interface and of buoyancy-driven mixing. The buoyancy-driven mixing experiment allows for long data collection time and is statistically steady. The experimental setup allows 1D mixing  simulations at the streamwise section. The two fluids have constant densities, $\rho_1 =$ 1.0833 g/cm$^3$   and $\rho_2 =$ 1.0 g/cm$^3$, to match the experiment. The buoyancy force is relatively weak  at this low Atwood number since it represents the effect of density difference on the flow mixture. The numerical results obtained with three different grid sizes are shown in Figure \ref{fig1} for normalized density, turbulent kinetic energy, and turbulent mass-flux velocity. The density distribution agrees very well with the experimental data \citep [] {Banerjee2010a}, and both the turbulent kinetic energy and mass-flux velocity show reasonable agreement. Furthermore, good grid convergence is obtained since the lines are on top of each other. Therefore, the following simulations with perturbations are performed using the $600$ grid size. 

\begin{figure}
\begin{center}
\includegraphics[width=1\textwidth]{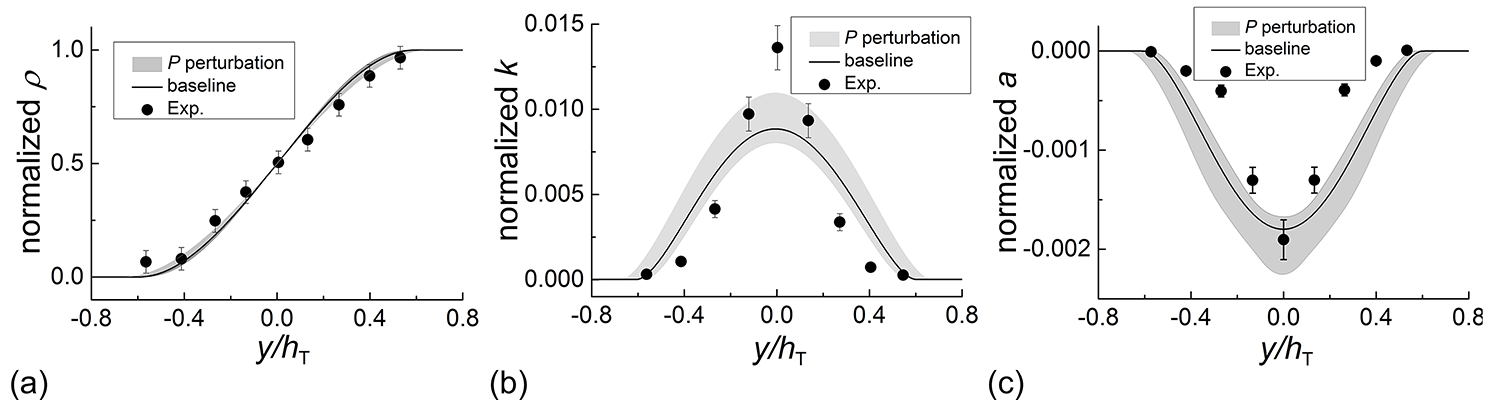}
\caption{Model uncertainty estimation quantified by the eigenspace perturbation in $P$ of 1DRT for (a) normalized density, (b) normalized turbulent kinetic energy, and (c) normalized turbulent mass-flux velocity.}
\label{fig2}
\end{center}
\end{figure}

Eigenspace perturbation in Reynolds stress is performed, and the five extreme cases \citep [] {Iaccarino2017} are simulated. The uncertainty bounds for the density, turbulent kinetic energy, and turbulent mass-flux velocity are presented in  Figure \ref{fig2}. As shown in  Figure \ref{fig1}(a), the numerical results agree very well with experimental data, and the uncertainty interval covers the experimental data. The uncertainty bounds of the turbulent kinetic energy and turbulent mass-flux velocity are able to account for the discrepancy between the model prediction and experimental data only near the center of the mixing layer. It seems that the five extreme eigenspace perturbation cases for the very simplified 1D case are insufficient to capture all of the model uncertainties in comparison to the complex three-dimensional experimental measurements.

\subsection{2DTR}

The TR problem is famous for a series of experiments \citep{Smeeton1987, Youngs1989} performed at Atomic Weapons Establishment (AWE) in the late 1980s. The light fluid is filled above the heavy one in a tank, and the apparatus is tilted to obtain an interface angle $\beta$.  Details of the configuration are presented by \cite {Andrews2014}. In this brief, we simply simulate experiment case 110 described by \cite{Smeeton1987} and \cite{Andrews2014}, which has an Atwood number of 0.48. The heavy fluid is a NaI solution with density, $\rho_1 =$ 1.89 g/cm$^3$, and the light fluid is hexane with density $\rho_1 =$ 0.66 g/cm$^3$. The interface angle $\beta = 5.7667$ deg in the rectangle rig with dimensions $L_x = 15$ cm and $L_y = 24$ cm. The tank acceleration for case 110 is time varying. However, there are two ways to represent the variable acceleration. We can incorporate the time-varying acceleration directly into the simulation, which is not convenient.  Alternatively, we can use a proper constant acceleration to nondimensionalize the time in order to compare with the experimental results. According to \cite{Andrews2014}, the constant acceleration $35 g$ agrees well with case 110. 

\begin{figure}
\begin{center}
\includegraphics[width=0.9\textwidth]{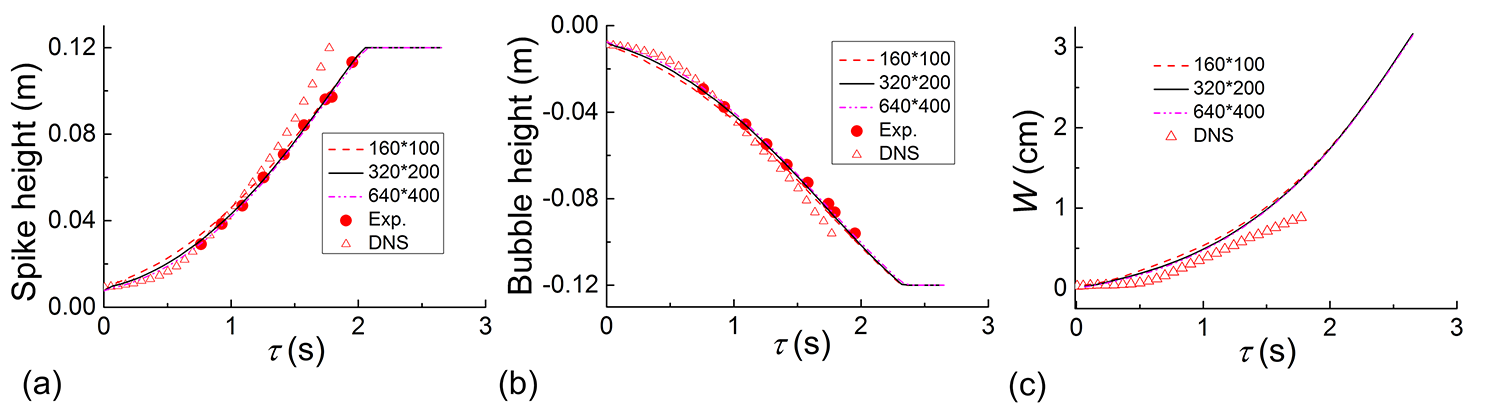}
\caption{ Simulations of 2DTR using the BHR model compared with experimental data \citep{Smeeton1987} and numerical results \citep{Andrews2014} for (a)  spike height, (b)  bubble height, and (c)  the mixing width.}
\label{fig3}
\end{center}
\end{figure}

2D simulations of buoyancy-driven flow are performed with three grid sizes, $160\times 100$, $320 \times 200$, and $640 \times 400$, and the results are presented in Figure \ref{fig3} comparing with the experimental data \citep{Smeeton1987} and numerical results \citep{Andrews2014}. All the numerical results of the three resolutions agree very well with the experiments and simulations. Very good grid convergence is obtained since the difference between the resolutions $320 \times 200$ and $640 \times 400$ is negligible. Therefore, grid size $320 \times 200$ is employed in the following simulations.

\begin{figure}
\begin{center}
\includegraphics[width=0.9\textwidth]{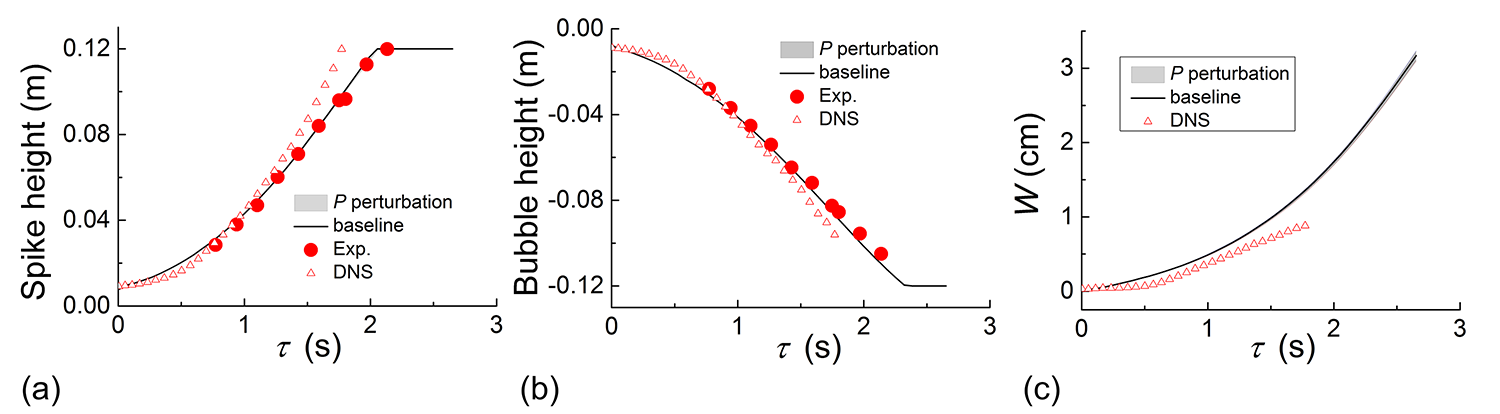}
\caption{Uncertainty estimation quantified by the eigenspace perturbation in $P$ of 2DTR for (a)  spike height, (b) bubble height, and (c) the mixing width.}
\label{fig4}
\end{center}
\end{figure}

We first performed the eigenspace perturbation in Reynolds stress. The uncertainty bounds for spike and bubble height and the mixing width are presented in Figure \ref{fig4}. The uncertainty intervals for all the quantities are negligible, which shows that the term $P$ in the turbulent kinetic energy transport equation should be very small for this buoyancy-driven flow. Therefore, the alignment or misalignment between the eigenvector of Reynolds stress and mean rate of strain is insufficient to affect the variable-density flow. 

\begin{figure}
\begin{center}
\includegraphics[width=0.9\textwidth]{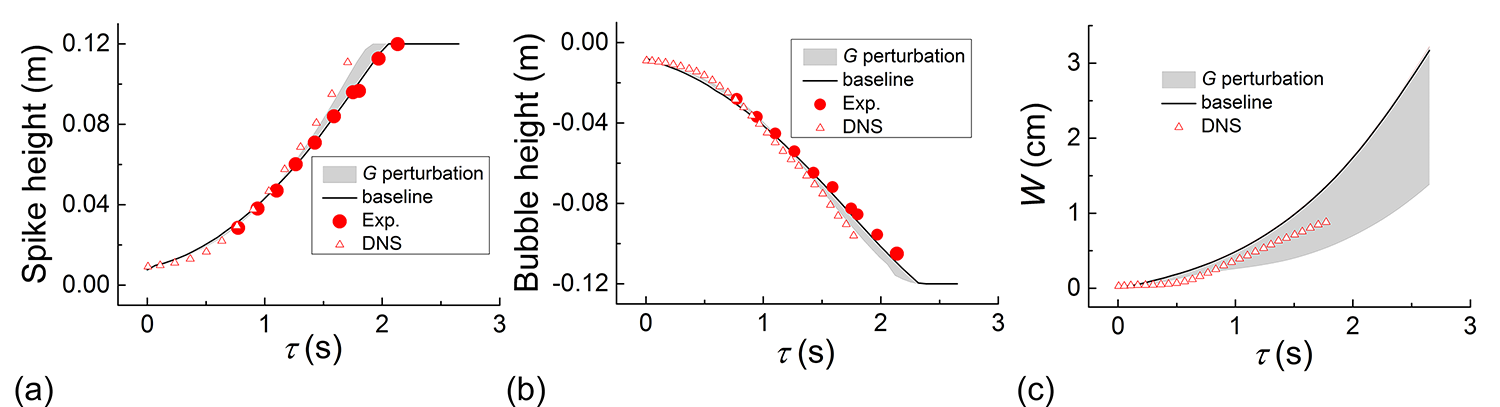}
\caption{Estimation of the model uncertainty considering the $G$ term of 2DTR for  (a) spike height, (b) bubble height, and (c)  the mixing width.}
\label{fig6}
\end{center}
\end{figure}

\begin{figure}
\begin{center}
\includegraphics[width=0.6\textwidth]{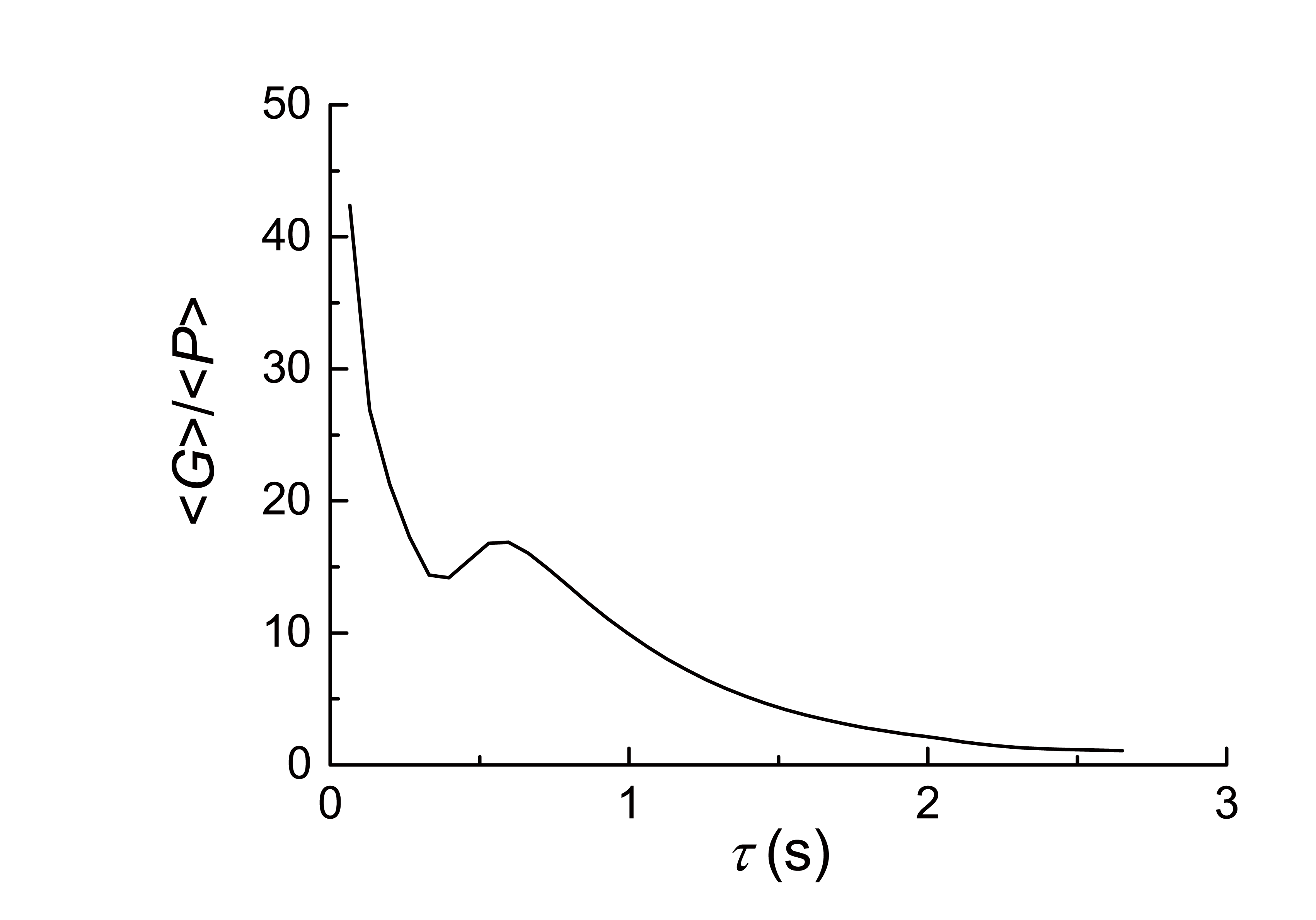}
\caption{The variation of the ratio between the volume integral of $G$ and $P$ with time.}
\label{fig5}
\end{center}
\end{figure}

 In the 2D set up, the angle between the vectors  $\bold{a}$  and  $\bold{p}_g$ could be varying from 0 to $\pi$. Since the flow is driven by buoyancy, the turbulent mass-flux velocity is most likely to obey the density gradient, or pressure gradient, and the angle between $\bold{a}$  and  $ \bold{p}_g$  varying from 0 to $\pi/2$ is more reasonable. Therefore, the term $G =  \bold{a} \cdot \bold{p}_g$ has the range $0 \le G \le \mid \bold{a} \mid \mid \bold{p}_g \mid$, where $\mid \mid$ is the magnitude of the vector.  Consequently, the two extreme cases of $G$ for buoyancy-driven flow are simulated, and the uncertainty estimation is shown in Figure \ref{fig6} in comparison to experimental data and DNS results. The uncertainty interval is significantly increased for $G$ perturbation compared with that of $P$ eigenspace perturbation, and enables us to account for the discrepancy between experimental observations and DNS results.
The ratio between the volume integral of $G$ and $P$, $ \langle G \rangle/\langle P\rangle$ where $\langle \psi \rangle  = \int_{V} \psi dV$ , is shown in Figure \ref{fig5}. It is obvious that the term $G$ is much larger than $P$, which explains the negligible uncertainty estimation of eigenspace perturbation in Reynolds stress shown in Figure \ref{fig4}.

\begin{figure}
\begin{center}
\includegraphics[width=0.9\textwidth]{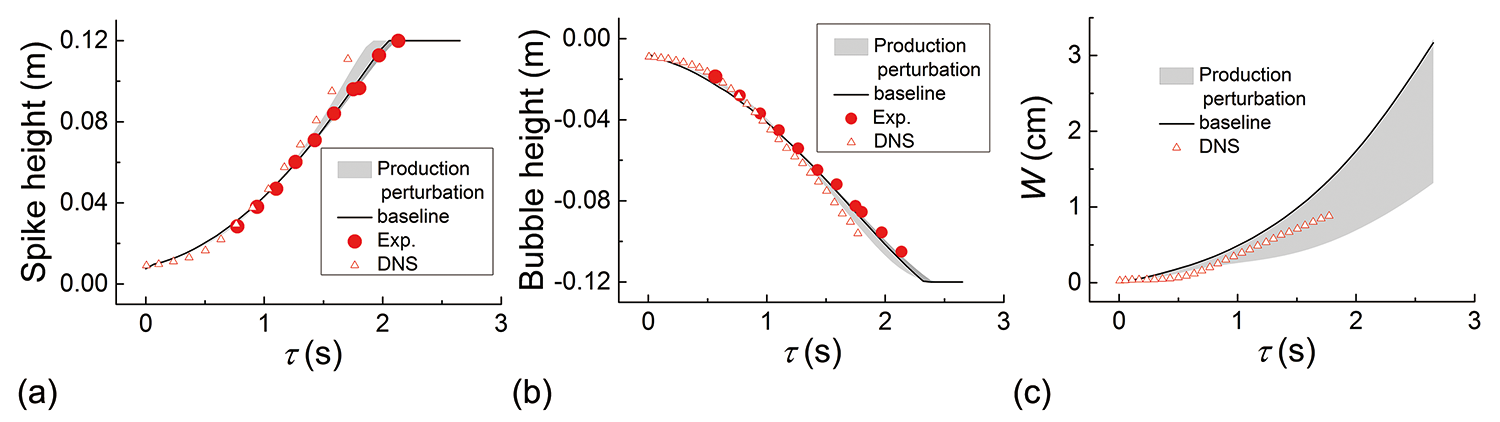}
\caption{The structural uncertainty quantified by the production perturbation of 2DTR for  (a) spike height, (b) bubble height, and (c)  the mixing width.}
\label{fig6p}
\end{center}
\end{figure}

Finally, the BHR model uncertainty is quantified by perturbing the two terms $G$ and $P$ together in the production. The uncertainty interval, shown in Figure \ref{fig6p}, is captured by the 10 extreme states.

\subsection{Turbulent jet mixing}

\begin{figure}
\begin{center}
\includegraphics[width=0.9\textwidth]{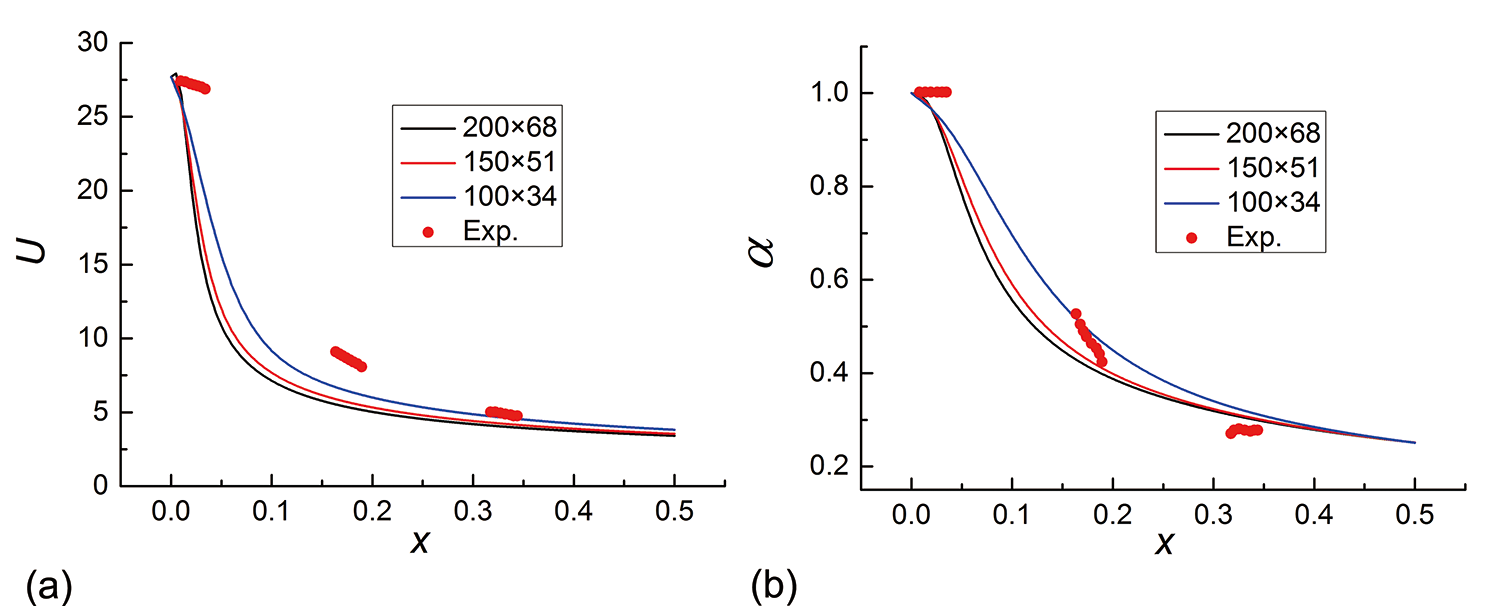}
\caption{Simulations of the turbulent jet mixing flow using the BHR model by comparison with experimental data \citep {Charonko2017} for (a) velocity and (b) mass fraction at the centerline.}
\label{fig7}
\end{center}
\end{figure}

In the above simulations, the flows are driven by buoyancy due to the variable density. Here, we consider the effect of density gradients on the transport of fluid and the evolution of the turbulent characteristics of the turbulent jet mixing flow. The turbulent jet flow has been studied experimentally by \citet {Charonko2017} using particle image velocimetry (PIV) and planar laser-induced fluorescence (PLIF) measurements. The jet flow is supplied using a $d_0 = 11$ mm inner-diameter tube vertically oriented in an open-circuit wind tunnel with a $0.53 \times 0.53$ m cross section. The acetone and gas mixture in the jet has density $\rho_1 = 1.1 $ g/cm$^3$, while the density of  ambient air  is $\rho_2 = 0.92 $ g/cm$^3$, which results in Atwood number $At = 0.09$. The inlet velocity of the jet is $27.7$ m/s, while the coflowing ambient air has velocity $1.4$ m/s. To simulate the flow, a jet with diameter $d_0 = 11$ mm in a circular channel with radius $r = 25 $ cm is chosen for the computational domain, the inlet turbulent kinatic energy of the jet $k = 20.0$ m$^2$/s$^2$ is selected. The calculated velocity and the mass fraction with different mesh sizes are compared with experimental data in Figure \ref {fig7}. The numerical results show reasonable agreement with experimental data, and the mesh size $150 \times 51$ is chosen for the following simulations. 

\begin{figure}
\begin{center}
\includegraphics[width=0.9\textwidth]{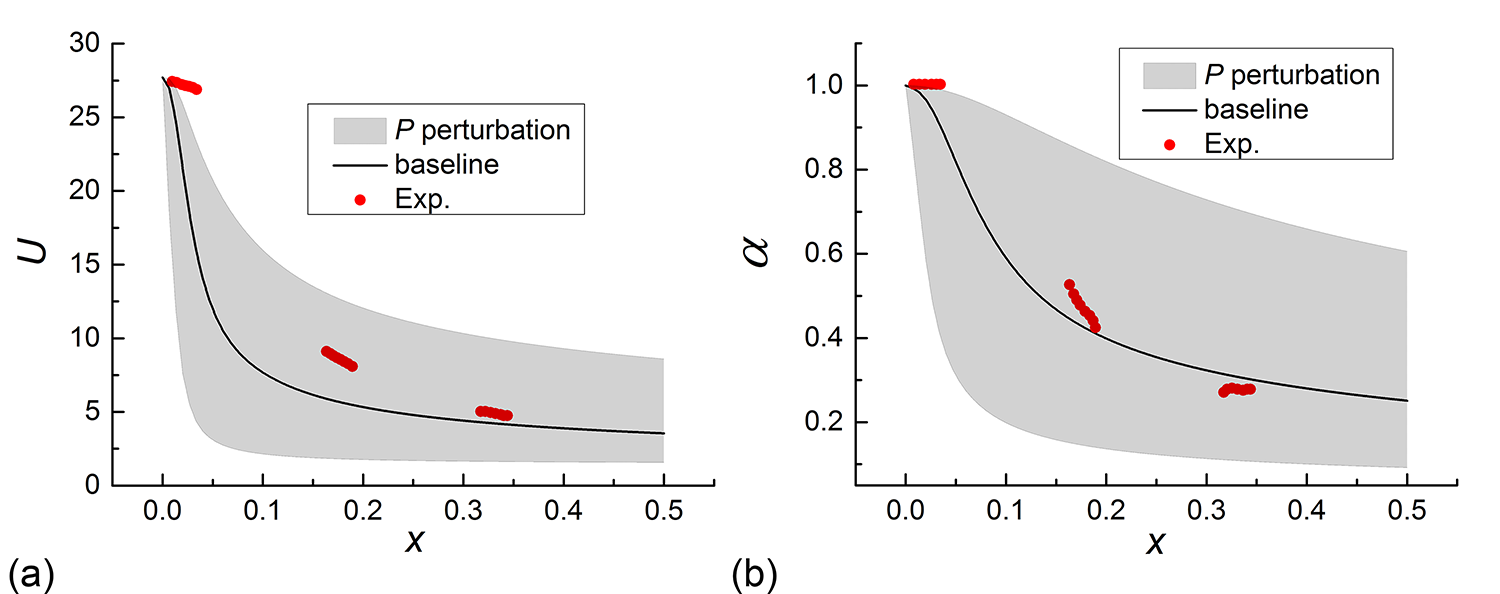}
\caption{Uncertainty estimation of the model by eigensapce perturbation in $P$ for turbulent jet mixing flow of  (a) velocity and (b) mass fraction at the centerline.}
\label{fig8}
\end{center}
\end{figure}

\begin{figure}
\begin{center}
\includegraphics[width=0.9\textwidth]{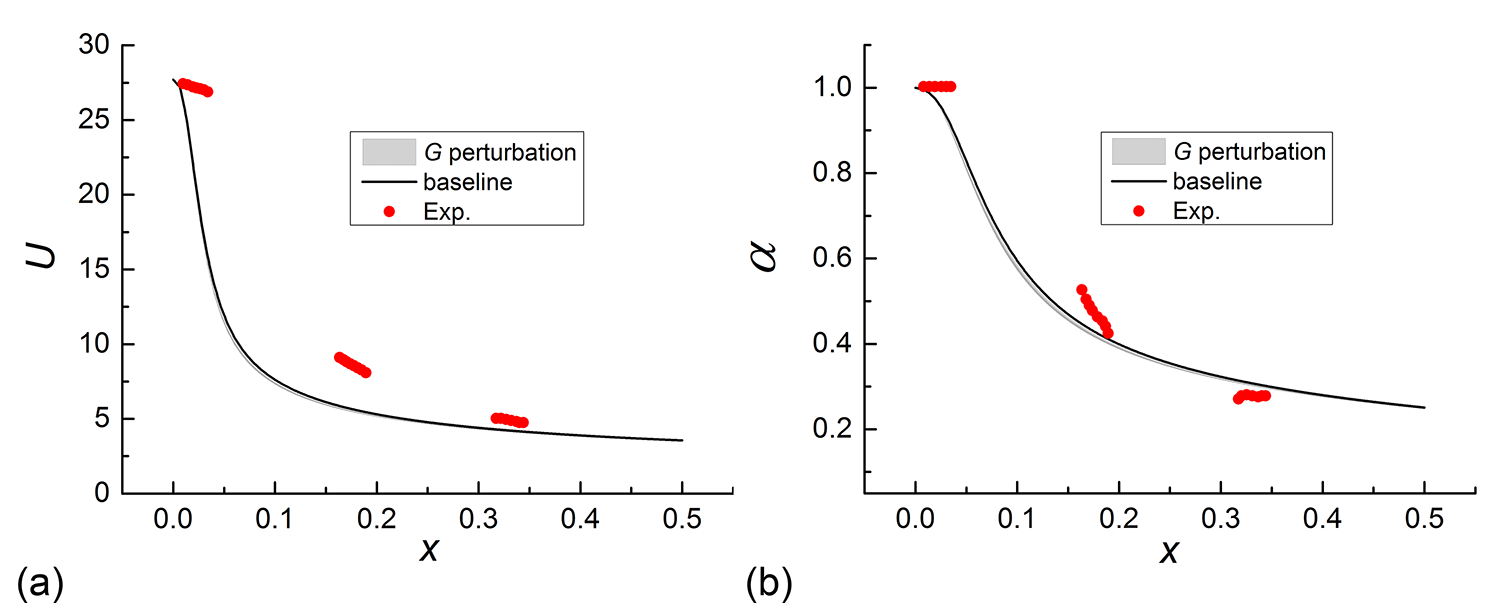}
\caption{Estimation of the uncertainty of the model by perturbing the $G$ term for turbulent jet mixing flow of (a) velocity and (b) mass fraction at the centerline.}
\label{fig8p}
\end{center}
\end{figure}

\begin{figure}
\begin{center}
\includegraphics[width=0.9\textwidth]{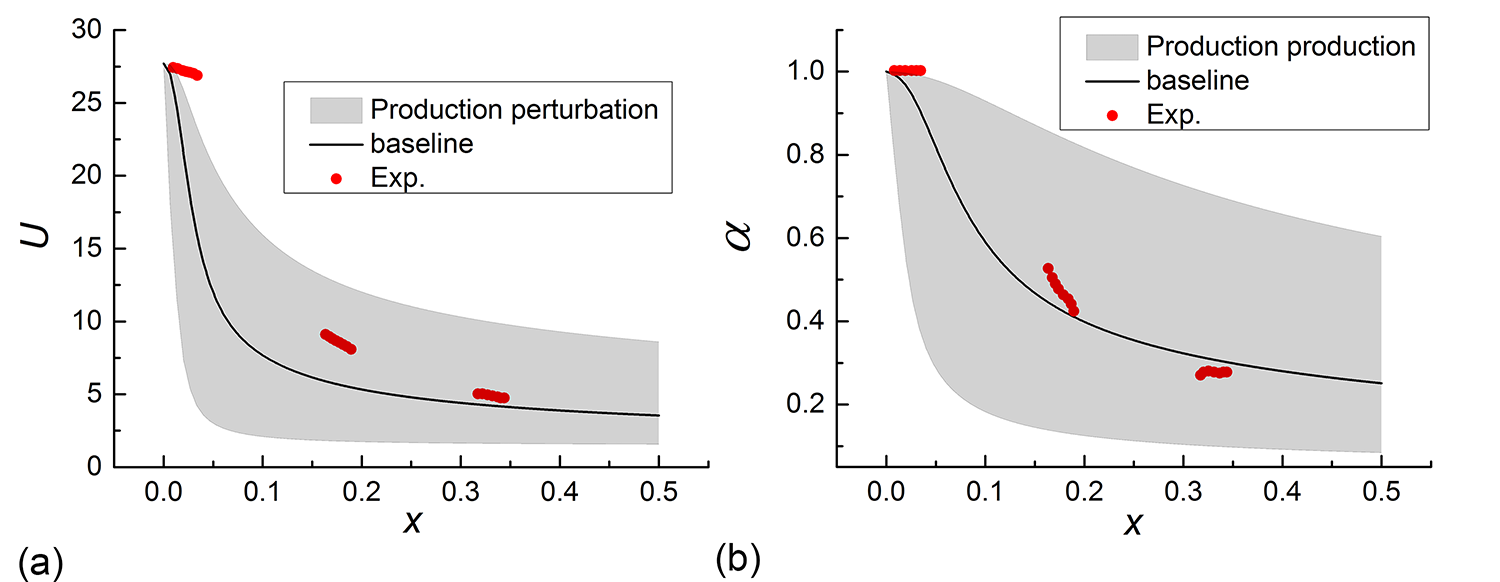}
\caption{Estimation of the model uncertainty by production perturbation for turbulent jet mixing flow of (a) velocity and (b) mass fraction at the centerline.}
\label{fig9}
\end{center}
\end{figure}

Unlike the TR case, the values of volume-averaged $P$ is about 24 times that of $G$ for the turbulent jet mixing flow, which means that $P$ is dominant. The uncertainty interval of eigenspace perturbation in $P$ is shown in Figure \ref {fig8}. The eigenspace perturbation bounds are able to account for the discrepancy between model predictions and experimental data except near the jet inlet, as expected since $P$ is dominant. In contrast, the uncertainty estimation of the $G$ term perturbation is negligible, as shown in  Figure \ref {fig8p}. 
The structural uncertainty quantified by maximizing and minimizing the production in the BHR model  is shown in  Figure \ref {fig9} by simulating the 10 extreme cases. Due to the combination of the eigenspace perturbation in $P$ and the perturbation in $G$, the perturbation of the production quantifies the structural uncertainty.

\section{Conclusions} 

Variable-density flows play an important role in a variety of physical phenomena and engineering applications. Simulations including DNS and LES are computationally expensive, especially for complex applications. Therefore, computationally tractable models for variable-density flows are more desirable in engineering applications. Therefore, the quantification of  structural uncertainty of the models are very important. In this brief, we have tried to build a comprehensive framework to identify, quantify, and reduce the uncertainties of the BHR model \citep {Besnard1992}, which have been extensively investigated for variable-density flows. In the BHR model, the production has two component terms, including the product of the Reynolds stress and rate of mean strain $ P$ and the inner product of the turbulent mass-flux velocity and pressure gradient $G$. By isolating the two terms, the effects of the eigenspace perturbation of $P$ and $G$ perturbation are discussed. According to our numerical results about 1DRT, Rayleigh-Taylor mixing in TR, and turbulent jet mixing flow, only consider one term in the production is not sufficient to capture the model uncertainty. The eigenspace perturbation in $P$ is not sufficient to account for the model uncertainty for $G$-dominant buoyancy-driven flows while $G$ perturbation is not sufficient to account for the model uncertainty for $P$-dominant variable density flows. Nevertheless, determining the the production perturbation by considering the $P$ and $G$ perturbation together enables us to quantify the structural uncertainty in the BHR model for variable-density flows. The proposed framework should be examined in more variable-density flows. 

\section*{Acknowledgments} 
This work is funded by a grant from Los Alamos National Laboratory.

\bibliographystyle{ctr}


\end{document}